\documentclass[a4paper,11pt]{article}
\pdfoutput=1

\usepackage{jcappub}

\usepackage{xcolor}
\usepackage{bm}
\usepackage{ulem}

\definecolor{grey}{rgb}{0.1,0.1,0.6}
\definecolor{green}{rgb}{0.1,0.5,0.4}

\title{\boldmath{The effective field theory approach to the strong coupling issue in $f(T)$ gravity}}

\author[a,b]{Yu-Min Hu,}
\author[a,b]{Yaqi Zhao,}
\author[a,b]{Xin Ren,}
\author[a,b]{Bo Wang,}
\author[c,a,d,1]{Emmanuel N. Saridakis,}
\author[a,b,1]{Yi-Fu Cai \note{Corresponding author.}}

\affiliation[a]{Deep Space Exploration Laboratory/School of  Physical Sciences, 
University of Science and Technology of China, Hefei, Anhui 230026, China}
\affiliation[b]{CAS Key Laboratory for Researches in Galaxies  and 
Cosmology/Department of Astronomy, School of Astronomy and Space Science, 
University of Science and Technology of China, Hefei, Anhui 230026, China}
\affiliation[c]{National Observatory of Athens, Lofos Nymfon, 11852 Athens, 
Greece}
\affiliation[d]{ Departamento de Matem\'{a}ticas, Universidad Cat\'{o}lica del 
Norte,  Avda. Angamos 0610, Casilla 1280 Antofagasta, Chile}

\emailAdd{yumin28@ustc.edu.cn}
\emailAdd{zxmyg86400@mail.ustc.edu.cn}
\emailAdd{rx76@mail.ustc.edu.cn}
\emailAdd{ymwangbo@ustc.edu.cn}
\emailAdd{msaridak@noa.gr}
\emailAdd{yifucai@ustc.edu.cn}

\abstract{ 
We investigate the scalar perturbations and the possible strong coupling issues of $f(T)$ around a cosmological background, applying the effective field theory (EFT) approach. We revisit the generalized EFT framework of modified teleparallel gravity, and apply it by considering  both linear and second-order perturbations for $f(T)$ theory. We find that no new scalar mode is present in both linear and second-order perturbations in $f(T)$ gravity, which suggests a strong coupling problem. However, based on the ratio of cubic to quadratic Lagrangians, we provide a simple estimation of the strong coupling scale, a result which shows that the strong coupling problem can be avoided at least for some modes. In conclusion, perturbation behaviors that at first appear problematic may not inevitably lead to a strong coupling problem, as long as the relevant scale is comparable with the cutoff scale $M$ of the applicability of the theory.  
}

\begin{document} 
\maketitle
\flushbottom

\section{Introduction}

The conception that gravity should be described through geometry was a revolution in gravitational physics, nevertheless by itself does not tell us what kind of geometry should be used. Despite the fact that in General Relativity (GR) by geometry one means Riemannian geometry, it was soon shown that this is not the only possibility \cite{CANTATA:2021ktz}. Maintaining simple geometries, one may have the geometric trinity of gravity 
\cite{BeltranJimenez:2019esp, Heisenberg:2018vsk}, in which gravitational interaction can be described with three different properties of spacetime geometry, namely  curvature $R$ for GR , torsion $T$ for teleparallel equivalent of general relativity (TEGR) \cite{Aldrovandi:2013wha, 
Maluf:2013gaa} and non-metricity $Q$ for symmetric teleparallel equivalent of general relativity (STEGR) \cite{Nester:1998mp}. 
Since $T$ or $Q$ are different from $R$ by a boundary term $B$ only, all three quantities when used as Lagrangians will lead to the same field equations, and thus the three formulations are equivalent. 

However, the above equivalence breaks when one proceeds to non-trivial modifications.   
The most simple and natural nonlinear extension of the TEGR action is to replace its Lagrangian $T$ with an arbitrary function $f(T)$, obtaining $f(T)$ modified gravity \cite{Cai:2015emx, Krssak:2018ywd}. 
Subsequently, one may construct a general framework which includes both $f(R)$ and $f(T)$ gravity, namely  $f(T,B)$ theory \cite{Bahamonde:2015zma, Bahamonde:2021gfp}. 
A lot of effort has been devoted to the investigation of the theoretical properties and observational implications of these theories \cite{Zheng:2010am, 
Bengochea:2010sg, Tamanini:2012hg, Cardone:2012xq, Farrugia:2016xcw, 
Bejarano:2017akj, Ong:2018heg, Chen:2019ftv, Cai:2019bdh,  Golovnev:2020nln, 
Golovnev:2020aon, 
Golovnev:2020las,Ren:2021uqb,BeltranJimenez:2021auj,Fiorini:2021mps,
Ren:2021tfi, Golovnev:2021htv,Mavromatos:2021hai,BeltranJimenez:2021kpj, 
Duchaniya:2022rqu, DeBenedictis:2022sja, Aljaf:2022fbk,
Zhao:2022gxl, Huang:2022slc,    
Jusufi:2022loj,   dosSantos:2021owt, Escamilla-Rivera:2019ulu, 
Caruana:2020szx, Bahamonde:2021srr, Moreira:2021xfe, Moreira:2021vcf, 
Shabbir:2020dbq, Moreira:2021cta, Najera:2022jvm, Malik:2021acs, 
Capozziello:2022dle, Kadam:2022lxt}, while it was shown that one can apply to them an effective field theory (EFT) approach, modifying and extending the original curvature-based EFT framework \cite{Li:2018ixg, Cai:2018rzd, Yan:2019gbw, Ren:2022aeo, Mylova:2022ljr}. 

Recently, there has been increasing interest in the physical degrees of freedom (DoFs)  that appear in  $f(T)$ gravity comparing to general relativity \cite{Diaz:2014yua,Diaz:2017tmy,Ferraro:2018axk, Ferraro:2020tqk, Golovnev:2020zpv}. 
Performing a Hamiltonian analysis, multiple studies \cite{Li:2011rn, Ferraro:2018tpu, Blagojevic:2020dyq,Blixt:2020ekl} have extracted  different results, yet all indicate there is at least one extra scalar-type DoF. 
Though all equations of $f(T)$ gravity are up to second order in derivatives, it doesn't guarantee healthy behavior. 
Nevertheless, a different path to study the dynamical modes of a theory on a given background is the perturbation analysis. 
Linear perturbations of $f(T)$ gravity in a Friendmann-Lema\^\i tre-Robertson-Walker (FLRW) universe have been investigated in  \cite{Dent:2010nbw, Chen:2010va, Izumi:2012qj, 
Golovnev:2018wbh, Sahlu:2019bug, Bahamonde:2022ohm, Hohmann:2020vcv}. 
As it turned out, in both background and linear perturbation level no extra DoFs appear. 
This point was further discussed in \cite{BeltranJimenez:2020fvy} around a Minkowski background, which manifested an extra dynamical scalar mode up to the fourth order perturbation action, and 
therefore the existence of the extra scalar-type DoF found through  perturbation analysis does not contradict with Hamiltonian analysis. 
However, the absence of this mode in lower order perturbations 
signals a possible strong coupling problem  in $f(T)$ gravity 
\cite{Bueno:2016xff, BeltranJimenez:2019tme, BeltranJimenez:2020lee}. 

Generally, in modified gravity,  a strong coupling problem usually refers to the phenomenon that the theory becomes strongly coupled at an extremely low energy scale \cite{Arkani-Hamed:2002bjr, Rubakov:2004eb, Deffayet:2005ys, 
Iglesias:2007nv, Charmousis:2009tc, Kimpton:2010xi, Clifton:2011jh, 
Papazoglou:2009fj, Blas:2009ck, Burrage:2012ja, Motohashi:2019ymr}. 
At the quantum level, the vacuum of a perturbative quantum field theory is not well defined \cite{Arkani-Hamed:2003pdi}, while at the classical level the predictions derived from the linear perturbation around a given background may not be convincing \cite{Baumann:2011dt}. 
Conditionally they can be healthy if a suitable nonlinear implementation is introduced, such as the Vainshtein mechanism \cite{Vainshtein:1972sx, Deffayet:2001uk}. Nevertheless, in  most of the situations it is not easy to find and apply one without conflicting with observations \cite{Dvali:2006su}.  

In order to examine a possible strong coupling problem in the case of $f(T)$ theory around a cosmological background, one natural way is to study the second-order perturbations of the theory and make an estimation of its strong-coupled scale. 
As long as it is high enough, strong coupling behavior would not appear at the relevant energy scales we are interested in. 
On the other hand, beyond this scale one may think that it 
should be described by an unknown ultraviolet-complete theory.

In this work, our main interest is to investigate cosmological perturbations of $f(T)$ gravity under the light of the possible strong coupling behavior, through the effective field theory  approach. Based on their EFT actions, we focus on the perturbations of scalar modes, and we study the dynamics in both quadratic and cubic order Lagrangians 
around a flat FLRW background.

This article is organized as follows. In Section \ref{sec: theory and EFT} we provide a short review of teleparallel gravity and its generalisation $f(T)$ gravity. 
In Section \ref{sec: EFT action}, we begin by reviewing a EFT action of a class of general torsional gravity, and apply to $f(T)$ gravity as well. 
Next, Section \ref{sec: scalar perturbation of f(T) gravity} focus on scalar perturbations of $f(T)$ gravity. 
We present the EFT action of $f(T)$ gravity up to cubic order, and investigate the scalar perturbations around a flat FLRW background in both quadratic and cubic action.
Then, in Section \ref{sec: strong coupling scale}, we offer an estimation on the strong coupling scale based on previous results.
Finally, the conclusions are provided in Section 
\ref{sec: conclusion}.

\section{Teleparallel gravity and the $f(T)$ generalisation}
\label{sec: theory and EFT}

In this section, we shortly review the foundations of  teleparallel geometry, and the basic forms of torsional action in the EFT approach. 
Teleparallel gravity is based on the tetrad field $e^{a}{}_{\mu}$ and the torsion tensor  ${T}^{\lambda}{}_{\mu \nu}$. 
The former is an orthonormal basis for the tangent space at each point $x^\mu$ of the manifold and is related to the metric through $g_{\mu \nu}=\eta_{a b} e^{a}{}_{\mu} e^{b}{}_{\nu}$, where Greek indices correspond to spacetime coordinates and Latin indices correspond to tangent space coordinates. 
In the following  we use the convention $\eta_{ab}=\text{diag}\{-,+,+,+\}$. 
The Weitzenb\"{o}ck connection  is built from the tetrad $e^{a}{}_{\mu}$ as: 
\begin{align}
\Gamma^\lambda{ }_{\nu \mu}=e_a{}^\lambda \partial_\mu e^a{ }_\nu~,
\end{align}
which leads to vanishing curvature corresponding to this connection.
Hence, the torsion tensor corresponding to the Weitzenb\"{o}ck connection is
\begin{align}
{T}^{\lambda}{}_{\mu \nu}=\Gamma^\lambda{ }_{\nu \mu}-\Gamma^\lambda{ }_{\mu 
\nu}=e_{a}{}^{\lambda}(\partial_{\mu} e^{a}{}_{ 
\nu}-\partial_{\nu} e^{a}{}_{\mu})~.
\end{align}
Moreover, using the relation between the metric and the tetrad, we can derive the relation between the Levi-Civita connection 
$\mathring{\Gamma}^{\rho}{}_{\mu \nu}$ in Riemannian spacetime and Weitzenb\"{o}ck connection in Weitzenb\"{o}ck spacetime, namely
\begin{equation}
\label{eq:conrelation}
	{\Gamma}^{\rho}{}_{\mu \nu}-\mathring{\Gamma}^{\rho}{}_{\mu 
\nu}={K}^{\rho}{}_{\mu \nu} ~,
\end{equation}
where the contortion tensor ${K}^{\rho}{}_{\mu \nu}$ is a combination of 
torsion 
 tensors, namely
$K^{\rho}{}_{\mu \nu} \equiv \frac{1}{2} \Big( T_{\mu}{}^{\rho}{}_{\nu} 
+{T_{\nu}{}^{\rho}}_{\mu} -{T^{\rho}}_{\mu \nu} \Big)$.
Through contractions of the torsion tensor one extracts the
torsion scalar \cite{Aldrovandi:2013wha, 
Maluf:2013gaa}:
\begin{equation}
    \label{eq:Tscalardef}
    T = S_{\rho}{}^{\mu \nu} T^{\rho }{}_{\mu \nu}=\frac{1}{4} T^{\rho}{}_{\mu 
\nu} T_{\rho}{}^{\mu \nu}+\frac{1}{2} T^{\rho}{}_{\mu \nu} T^{\nu 
\mu}{}_{\rho}-T^{\rho}{}_{\mu \rho} T^{\nu \mu}{}_{\nu} ~,
\end{equation}
where $
	S_{\rho}{}^{\mu \nu} \equiv \frac{1}{2} \Big( {K}^{\mu\nu}{}_{\rho} 
+\delta_{\rho}^{\mu} T^{\alpha \nu}{}_{\alpha} -\delta_{\rho}^{\nu} T^{\alpha 
\mu}{}_{\alpha} \Big)$ is the super-potential.

According to \eqref{eq:conrelation} and (\ref{eq:Tscalardef}), as well as the definition of the Ricci scalar $R$ corresponding to the Levi-Civita connection, 
the relation between $R$ and  $T$ can be extracted, namely 
\cite{Aldrovandi:2013wha, Maluf:2013gaa}
\begin{equation}
R=-T-2 \nabla_\mu T^\mu,
\label{eq:relationRT}
\end{equation}
where $T^\mu=T^{\nu \mu  }{}_{\nu}$ is the contraction of the torsion tensor, and the ``boundary term'' is defined as $B\equiv -2 \nabla_\mu T^\mu$. 
As is shown above, since $R$ and $T$ differ only by a boundary term, GR and TEGR will have the same equations, which indicates that both are equivalent in this level.

Inspired by the  $f(R)$ modifications of gravity, one can proceed to the construction of modified teleparallel gravity by upgrading the torsion scalar in the action to a function $f(T)$, resulting to $f(T)$  gravity, namely
\begin{equation}
 S = \int d^{4} x\, e \,\frac{M_{P}^{2}}{2} f({T}) ~,
\end{equation}
where the torsion scalar $T$ is constructed by the Weitzenb\"{o}ck connection, $e=\det{(e{^a}{_\mu})=\sqrt{-g}}$ and $M_{P}$ is the Planck mass. 
However, since in general a function of a boundary term is not a boundary term,   $f(R)$ gravity and $f(T)$ gravity will lead to different field equations and thus they are different theories. 
One can go further and write the action as a function of the torsion scalar $T$ and the boundary term $B$, which is the $f(T,B)$ gravity \cite{Bahamonde:2015zma, Bahamonde:2021gfp}. 
As we can see, due to  \eqref{eq:relationRT} we have $f(R)=f(-T+B)$, and thus  $f(R)$ gravity is a subclass of $f(T,B)$ theory.

\section{ The EFT action of  $f(T)$ gravity} 
\label{sec: EFT action}

In this section, we will briefly introduce the EFT approach of torsional gravity. 
Taylor expansion will also be applied to $f(T)$ gravity to obtain its corresponding EFT form.

As we mentioned in the Introduction, in the present work we mainly focus on cosmological perturbations and the possible strong coupling behaviors. 
For this purpose, we will apply the EFT approach \cite{Arkani-Hamed:2003pdi, Cheung:2007st, Creminelli:2008wc, Gubitosi:2012hu, Gleyzes:2013ooa, Piazza:2013coa, Ashoorioon:2018uey}, since it has been shown to exhibit many advantages. 
On the one hand, EFT formalism provides significant capabilities, allowing for model-independent constraints \cite{Bloomfield:2012ff}, since its application allows one to investigate the background and perturbations separately by an expansion parameter up to a given order (note that different theories may share a common EFT formulation in a simple form at the background level).
On the other hand, EFT is a powerful tool dealing efficiently with the DoFs of the theory, in the form of perturbations, at the relevant energy scales. 
In other words, it is convenient to select the leading terms and investigate the strong coupling behavior between linear and nonlinear perturbations. 
Finally, in this framework the strong coupling scale can be viewed as a cutoff of the low-energy approximation \cite{Blas:2009ck}. 

The general EFT action of general torsional gravity is given by 
\cite{Li:2018ixg}
\begin{align}
 S = \int d^4x \sqrt{-g} \Big[ \frac{M^2_P}{2} \Psi(t)R - \varLambda(t) - b(t) 
g^{00} + \frac{M^2_P}{2} d(t) T^0 \Big] +S^{(2)}+\cdots ~,
 \label{generalEFT}
\end{align}
where  $\Psi(t)$, $\varLambda(t)$, $b(t)$ and $d(t)$ are time-dependent coefficients. 
All terms that explicitly involve second-order perturbations as their leading contribution are contained in $S^{(2)}$, while the dots denote higher-order perturbations.
Since the unitary gauge has been imposed, the time coordinate is fixed, which leads to a breaking of time diffeomorphism  invariance. 
Thus, the EFT formulation is constructed by operators invariant under spatial diffeomorphisms. 
Moreover, the contracted torsion tensor $T^0$ is contained as a basic operator reflecting the nature of torsional geometry  in the EFT language. 
Note that the Ricci scalar $R$ may also appear in the pure torsional part through relation \eqref{eq:relationRT}, namely the torsion scalar $T$  with the ``boundary term'' $T^0$. 
After restoring the diffeomorphism symmetry applying the St\"{u}ckelberg trick, the Nambu–Goldstone mode $\pi$ is present, and the additional scalar-type DoF becomes explicit and mixing in this representation. 
Finally, the corresponding Friedmann equations around a flat FLRW geometry are obtained through variation, and govern the evolution of the background \cite{Li:2018ixg}.

The EFT action of $f(T)$ gravity is constructed by expanding 
\begin{align}
f\left(T\right)=f(T^{(0)})+f_{T}(T^{(0)})\left(T-T^{(0)}\right)+\frac{1}{2}f_{TT}(T^{(0)})\left(T-T^{(0)}\right)^{2}+\cdots~,
\end{align}
where  subscript $T$ denotes the derivative with respect to the torsion scalar 
$T$, i.e. $f_T=\frac{df}{dT}$. In this section, we don't impose any constraint on the functional form of $f(T)$. 
Thus, it is easy to return to TEGR case, namely GR counterpart, in the appropriate limit. 
Additionally, $T^{(0)}$ is used to represent the torsion scalar at the background level with the value $T^{(0)}=6H^{2}$ (not to be confused with the zero-th component of the contracted torsion tensor $T^0$). Hence the EFT action at leading order is found to be
\cite{Li:2018ixg,Mylova:2022ljr}
\begin{align}
\label{EFTactionofFT}
S=\int d^{4}x\sqrt{-g} 
\left\{\frac{M_{P}^{2}}{2}\Big[f_{T}(T^{(0)})T+f(T^{(0)})-f_{T}(T^{(0)})T^{(0)}
\Big]-\varLambda(t) \right\},
\end{align}
where we have  focused on the leading-order operators, and we have added a cosmological constant term for completeness.
Replacing the torsion scalar $T$ through  relation \eqref{eq:relationRT}, we can rewrite the action in an equivalent form as 
\begin{align}
S=\int \!
d^{4}x\sqrt{-g}\left\{\frac{M_{P}^{2}}{2}\Big[\!-f_{T}(T^{(0)})R+2\dot{f}_{T}(T^
{ (0)})T^{0}-f_{T}(T^{(0)})T^{(0)}+f(T^{(0)})\Big]-\varLambda(t) \!
\right\}.
\label{fTleading}
\end{align}
Note that for simplicity we have expanded all the coefficient functions as ordinary time-dependent functions, namely $f=f(t)$, $~f_{T}=f_{T}(t)$ and $\dot{f}_{T}=\dot{f}_{T}(t)$, and these functions are assumed to be non-zero in general.

\section{Scalar perturbations of $f(T)$ gravity}
\label{sec: scalar perturbation of f(T) gravity}

\subsection{Background evolution}

Let us first examine the background evolution. Since in torsional geometry the dynamical quantities are the tetrad fields, we consider a flat FLRW geometry in the mostly negative signature, which arises from the tetrad
\begin{eqnarray}  
e^0{}_{\mu} &=& \delta^0_{\mu}  ~,  \\  
e^a{}_{\mu} &=& a\delta^i_{\mu} \delta^a_i  ~. 
\label{basictetrad00}
\end{eqnarray}
This tetrad choice is commonly used in the literature, although not unique. Inserting this into  (\ref{eq: EFT for f(T)})
and performing variation, we extract the field equations
\begin{align}
\label{eq: EoM b}
\varLambda(t) &=\frac{1}{2}M_{P}^{2}\left(f-12H^{2}f_{T}\right)~, \\
\label{fTdotH}
f_{T}\dot{H} &=- H\dot{f}_{T} ~, 
\end{align}
where we have used the background value 
$T^{(0)}=6H^{2}$. Note that we can compare  \eqref{eq: EoM b} 
and \eqref{fTdotH} with the background equations (4.1) and (4.2) in 
\cite{Li:2018ixg}, which (neglecting the matter part) take the form   
\begin{align}
    \label{eq: Friedmann1}
    -M_{P}^{2}f_{T} \Big( -\dot{H}-H\frac{\dot{f}_{T}}{f_{T}} \Big)	
&
    =0~, 
    \\
    \label{eq: Friedmann2}
    -3M_{P}^{2}f_{T}H^{2}-\frac{1}{2}M_{P}^{2} \big[T^{(0)}f_{T}-f\big]	
&
    =\varLambda~.
\end{align} 
 As we can see, \eqref{eq: EoM b} and \eqref{fTdotH} are consistent with 
\eqref{eq: Friedmann1} and \eqref{eq: Friedmann2}, and both can be regarded as the Friedmann equations of $f(T)$ gravity with a time-dependent function term $\varLambda(t)$.
In this sense, both equations don't constraint the function form of $f(T)$. Then, as we mentioned previously, GR would be easily recovered as a special case.

\subsection{Scalar perturbations of the tetrads} 
 
In the following we will examine the scalar perturbations. The perturbed tetrads up to linear order are written in the Newtonian gauge and take the form
\begin{eqnarray}  
e^0{}_{\mu} &=& \delta^0_{\mu} +\delta^0_{\mu}\phi +a\delta^i_{\mu} 
\partial_i\chi ~,  \\  
e^a{}_{\mu} &=& a\delta^i_{\mu} \delta^a_i(1-\psi) +\delta^0_{\mu}\delta^a_i 
\partial^i\chi~,
\label{basictetrad}
\end{eqnarray}
with $\phi$, $\psi$, $\chi$ the scalar degrees of freedom, and where we have neglected the  pseudoscalar  to avoid parity violation. 
Using
$e_{~\mu}^{a}=\delta_{~\nu}^{a}e_{~\mu}^{\nu}$,  where
 $\delta e_{~\mu}^{\nu} $ is the perturbation, we have  \cite{Li:2018ixg}
\begin{eqnarray}
e_{~\mu}^{\nu}= \delta_{~\mu}^{\nu}+\delta 
e_{~\mu}^{\nu}+\frac{1}{2}\delta e_{~\rho}^{\nu}\delta e_{~\mu}^{\rho}+\cdots~ ,
\label{Eexpand}
\end{eqnarray}
up to the desired  order.
For instance, the perturbed tetrads up to second order are expressed as:
\begin{align}   	  e^0_{\mu} =& \delta^0_{\mu} \big( 1 +\phi +\frac12\phi^2 
+\frac12\partial_i\chi \partial_i\chi \big) + a\delta^i_{\mu} \Big[ \partial_i 
\chi + \frac12(\phi\partial_i\chi -\psi\partial_i\chi) \Big] ~, \\  e^a_{\mu} =& 
a\delta^i_{\mu} \delta^a_i \big( 1-\psi+\frac12\psi^2 \big) + \frac{a}{2} 
\delta^i_{\mu} \delta^a_j \partial_i \chi \partial_j \chi + \delta^0_{\mu} 
\delta^a_i \Big[ \partial_i\chi +\frac12(\phi\partial_i\chi - 
\psi\partial_i\chi) \Big] ~, \\  e^{\mu}_0 =& \delta^{\mu}_0 \big( 1 -\phi 
+\frac12\phi^2 +\frac12\partial_i \chi\partial_i\chi \big) + 
\frac1a\delta_i^{\mu} \Big[ -\partial_i\chi +\frac12(\phi\partial_i\chi - 
\psi\partial_i\chi) \Big] ~, \\  e^{\mu}_a =& \frac1a \delta_i^{\mu} \delta_a^i 
\big( 1 + \psi + \frac12 \psi^2 \big) + \frac{1}{2a} \delta_i^{\mu} \delta_a^j 
\partial_i \chi \partial_j \chi + \delta_0^{\mu} \delta_a^i \Big[ 
-\partial_i\chi +\frac12(\phi\partial_i\chi - \psi \partial_i \chi) \Big] ~. 
\end{align} 
Then the perturbed metric components up to second order are accordingly given 
by 
\begin{align}  
g_{00} &= -(1+2\phi+2\phi^2) ~, \\ 
g_{0j} &=-a(\phi+\psi)\partial_j \chi ~,\\
g_{ij} &= a^2\delta_{ij}(1-2\psi+2\psi^2) ~, \\  
g^{00} &= -(1-2\phi+2\phi^2) ~, \\ 
g^{0j} &=-a^{-1}(\phi+\psi)\partial^{j} \chi ~,\\
g^{ij} &= a^{-2}\delta_{ij}(1+2\psi+2\psi^2) ~, 
\end{align}
while the determinant of the metric is
\begin{align}
 \sqrt{-g} = a^3 \left(1 -3\psi +\phi +\frac{9}{2}\psi^2 -3\phi\psi 
+\frac12\phi^2\right) ~.
\end{align}

We can now apply the  St\"{u}ckelberg trick, and thus we can restore the general coordinate invariance of the theory. 
After performing a time coordinate transformation  of the form $t\rightarrow t+\pi$ on the unitary gauge action, the Nambu–Goldstone mode $\pi$ would appear in the action. 
Additionally,  $T^{(0)}$ should also be regarded as a time-dependent function and expanded as
\begin{align}
T^{(0)}\rightarrow  T^{(0)}+\dot{T}^{(0)}\pi+\frac12\ddot{T}^{(0)}\pi^2+\cdots~.
\end{align}
Finally, for the time-component of the contracted torsion $T^{0}$  we have
\begin{align}
T^{0}\rightarrow T^{0}+\partial_{\mu}\pi\,T^{\mu}~.
\end{align}

\subsection{Linear-order scalar perturbations of $f(T)$ gravity}
In this section  we investigate linear scalar perturbations in the case of $f(T)$ gravity around a flat FLRW background in the absence of matter couplings. 

Inserting the perturbed tetrads into  the leading order EFT form (\ref{fTleading}),
we obtain the kinetic part of the quadratic Lagrangian, i.e. which includes second-order and higher-order spacetime derivatives, as
\begin{align}
\mathcal{L}_{2}^{kin}=M_{P}^{2}a\left\{f_{T}\left[3a^{2}\dot{\psi}
^2+\partial^i\psi\left(2\partial_i\phi-\partial_i\psi\right)\right]+2a\dot{f}_{T
}\left(H\partial_i\pi+\partial_i\psi\right)\partial^i\chi\right\}~,
\end{align} 
\label{eq: L2kin}
which is exactly the same with the result of Ref \cite{Li:2018ixg}. 
However, we mention that in that manuscript the authors worked in Fourier space, and focused on modes that are deep inside the horizon. Based on such consideration, the dispersion relation $f_{T}^{2}\dot{f}_{T}^{2}k^{8}=0$ is 
obtained by $\mathcal{L}_{2}^{kin}$, which implies that there are no propagating scalar modes in $f(T)$ gravity at this level. 
Finally, as expected for consistency, if the condition $\dot{f}_{T}=0$ is satisfied  the model  returns to  TEGR. 

In order to perform a complete analysis on linear perturbation level, both linear and second-order operators are considered in this subsection. 
Because of the lack of a general torsional EFT action up to second order, especially the lack of a complete set of second-order operators), we just focus on this simple action: 
\begin{equation}
\label{eq: EFT for f(T)}
S=\int d^{4}x\sqrt{-g}\left\{ \frac{M_{P}^{2}}{2}\Big[f(T^{(0)})+f_{T}(T^{(0)})\left(T-T^{(0)}\right)+\frac{1}{2}f_{TT}(T^{(0)})\left(T-T^{(0)}\right)^{2}\Big]-\varLambda(t)\right\} ~,
\end{equation}
without rewriting it in a standard and simplified EFT form, and investigate the perturbation behaviour of this form.  
Then plugging the perturbed tetrads into \eqref{eq: EFT for f(T)} we acquire
\begin{align}
\mathcal{L}_{2}^{kin}= & M_{p}{}^{2}a\left(f_{T}\left(3a^{2}\dot{\psi}^{2}+\partial_{i}\psi\left(2\partial^{i}\phi-4aH\partial^{i}\chi-\partial^{i}\psi\right)\right)\right.\\ \nonumber
 & \left.+2aH\left(6f_{TT}H\left(3a\dot{\psi}^{2}-2H\partial_{i}\chi\partial^{i}\phi\right)+\dot{f_{T}}\partial_{i}\chi\partial^{i}\pi\right)\right)\:.
\end{align}

As it is known,  theories with multiple fields in general allow for 
a much richer constraint structure, which is highly complicated since these 
fields are kinetically mixed with each other. Thus, the behavior of the 
propagating mode, arising from the specific form of the Lagrangian, can be 
affected   by the non-dynamical degrees of freedom, too. We refer to 
Refs. \cite{Gracia:1988xp,Henneaux:1990au} for the constraint analysis and DoF 
counting in the Lagrangian approach is various such  cases. 

In our EFT formulation, the quadratic perturbative action contains four 
different scalar fields, namely $\psi$, $\pi$, $\chi$ and $\phi$. According to 
the specific form of $\mathcal{L}_{2}^{kin}$ in  \eqref{eq: L2kin}, $\pi$, 
$\chi$ and $\phi$ act as auxiliary variables, i.e. they do not have  
time derivatives. We emphasize that the existence of auxiliary variables is 
crucial in imposing the constraints and avoiding the propagation of the extra 
mode \cite{deRham:2016wji,Gao:2019lpz,Hu:2021yaq}. In order to reduce the 
complications resulting from the presence of multiple fields, we reduce the 
number of scalar fields by solving these constraints. This method can be applied 
in linear perturbation analysis, and the solutions would applied in second-order 
perturbations, which is one of the central topics of this work.

For these reasons, before investigating the dynamics of the theory, we 
first perform   variations with respect to each variable. The constraint equation derived from the variation with respect to $\pi$ and $\phi$ are 
\begin{align}
\partial^{2}\chi= & -3a\left(H\phi+\dot{\psi}-H\frac{\dot{f}_{T}}{f_{T}}\pi\right)~,\\
\phi= & \frac{1}{3a^{2}f_{T}H^{2}\left(f_{T}+12f_{TT}H^{2}\right)}\\ \nonumber
 & \times\Big[f_{T}^{2}(\partial^{2}\psi-3a^{2}H\dot{\psi})+36a^{2}f_{TT}H^{4}\pi\dot{f}_{T}\\ \nonumber
 & \quad-3af_{T}H^{2}\left(4f_{TT}H(\partial^{2}\chi+3a\dot{\psi})-a\pi\dot{f}_{T}\right)\Big] \nonumber
\end{align}
After plugging the solution and make simplification, we
have 
\begin{align}
\label{fTchi} 
\chi & =-\frac{\psi}{aH}~, \\ 
\label{fTphi}
\phi & =\frac{\partial^{2}\psi}{3a^{2}H^{2}}-\frac{\dot{\psi}}{H}+\frac{\dot{f}_{T}}{f_{T}}\pi~. 
\end{align}
Making use of both solution, we simplified the other constraint equation
\begin{align} 
2M_{P}^{2}a	& \Big[-12a\frac{f_{TT}}{f_{T}}\dot{f}_{T}H^{3}\partial^{2}\pi-a\dot{f}_{T}\left(H\partial^{2}\pi+\partial^{2}\psi\right) \\ \nonumber
&	\quad +4f_{TT}H^{2}\left(3aH\partial^{2}\phi+\partial^{4}\chi+3a\partial^{2}\dot{\psi}\right)\Big]=0
\end{align}
as
\begin{align} 
-2a^{2}\left(H\partial^{2}\pi+\partial^{2}\psi\right)\dot{f}_{T}=0~.
\end{align}
This constraint is automatically satisfied if $\dot{f}_{T}=0$, i.e.
in the TEGR case. Assuming that $\dot{f}_{T}\neq0$, we solve for
the non-dynamical field $\pi$ as 
\begin{equation}
\pi=-\frac{\psi}{H}~.\label{fTpi}
\end{equation}
Hence, as we observe from \eqref{fTpi} and \eqref{fTchi}, 
 $\pi$  and $\chi$   are not relevant in the subsequent dynamical analysis of 
second-order perturbations.

Substituting the above solutions into the quadratic Lagrangian of $f(T)$ theory and performing integration by parts, we can rewrite it as
\begin{align}
\mathcal{L}_{2}=-M_{P}^{2}\frac{1}{3aH^{2}}f_{T}\left(\partial^{2}\psi\right)^{2}~.
\label{fTk4}
\end{align} 
This equivalent formulation of the quadratic Lagrangian has no kinetic time 
derivatives of the involved modes. Hence, we conclude that  no dynamical 
scalar mode exists in linear perturbation of $f(T)$ theory. This implies that  
 $f(T)$ theory may suffer from a strong coupling problem around 
the cosmological background. That is why, in order to examine this issue more 
deeply, in the next subsection we proceed to 
the  second-order scalar 
perturbations.

\subsection{Second-order scalar perturbations of $f(T)$ gravity}
In this subsection we proceed to second-order perturbation analysis. We start by perturbing the tetrads up to cubic order, in which case expansion 
\eqref{Eexpand} has the additional contribution:
\begin{align}   	 
~^{(3)}\delta e^0_{\mu} =& 
\frac{1}{6}\delta_{\mu}^{0}\left[\phi^{3}+(2\phi-\psi)\,(\partial_i\chi)^{2}
\right]+\frac{1}{6}a\delta_{\mu}^{i}\partial_{i}\chi\left[\phi^{2}
-\phi\psi+\psi^{2}+(\partial_j\chi)^{2}\right]~, \\  
~^{(3)}\delta e^a_{\mu} =&  
-\frac{1}{6}a\delta_{\mu}^{i}\delta_{i}^{a}\psi^{3}+\frac{a}{6}\delta_
{\mu}^{i}\delta_{j}^{a}(\phi-2\psi)\partial_{i}\chi\partial_{j}\chi+\frac{1}{6}\delta_{\mu}^
{0}\delta_{i}^{a}\partial_{i}\chi\Big[\phi^{2}-\phi\psi+\psi^{2}
+(\partial_j\chi)^{2}\Big]~, \\  
~^{(3)}\delta e^{\mu}_0 =& 
\frac{1}{6}\delta_{0}^{\mu}\left[-\phi^{3}+(-2\phi+\psi)\,(\partial_i\chi)^{2}
\right]-\frac{1}{6a}\delta_{i}^{\mu}\partial_{i}\chi\left[\phi^{2}
-\phi\psi+\psi^{2}+(\partial_j\chi)^{2}\right] ~, \\  
~^{(3)}\delta e^{\mu}_a =&  
\frac{1}{6a}\delta_{i}^{\mu}\delta_{a}^{i}\psi^{3}-\frac{1}{6a}\delta_{
i}^{\mu}\delta_{a}^{j}(\phi-2\psi)\partial_{i}\chi\partial_{j}\chi-\frac{1}{6}
\delta_{0}^{\mu}\delta_{a}^{i}\partial_{i}\chi\Big[\phi^{2}-\phi\psi+\psi^{2}
+(\partial_j\chi)^{2}\Big]~, 
\end{align} 
which gives rise to the  cubic-order  metric perturbation:
\begin{align}  
~^{(3)}\delta g_{00} &= 
-\frac{2}{3}\left[2\phi^{3}+(\phi+\psi)\,(\partial_i\chi)^{2}\right] ~, \\  
~^{(3)}\delta g_{0j} &=-a (\phi^2-\psi^2)\partial_j\chi
~, \\ 
~^{(3)}\delta g_{ij} &= 
-\frac{2}{3}a^{2}\left(2\delta_{ij}\psi^{3}+(\phi+\psi)\,\partial_{i}
\chi\partial_{j}\chi\right) ~, \\  
~^{(3)}\delta g^{00} &= 
\frac{2}{3}\left[2\phi^{3}+(\phi+\psi)\,(\partial_i\chi)^{2}\right] ~, \\  
~^{(3)}\delta g^{0j} &=\frac{1}{a} (\phi^2-\psi^2)\partial^j\chi
~, \\ 
~^{(3)}\delta g^{ij} &= 
\frac{2}{3a^{2}}\left(2\delta^{ij}\psi^{3}+(\phi+\psi)\,\partial^{i}
\chi\partial^{j}\chi\right) ~, 
\end{align}
which implies the determinant 
\begin{align}
\label{g}
 \sqrt{-g} = a^3 \left[1 +\phi-3\psi+\frac12 (\phi -3\psi )^2 +\frac16 (\phi 
-3\psi )^3 \right] ~.
\end{align}
Similarly, we expand the function form of $f(T)$ up to cubic order as
\begin{align}  
f\left(T\right)=& f(T^{(0)})+f_{T}(T^{(0)})\left(T-T^{(0)}\right)+\frac{1}{2}f_{TT}(T^{(0)})\left(T-T^{(0)}\right)^{2}\nonumber
\\
&+\frac{1}{6}f_{TTT}(T^{(0)})\left(T-T^{(0)}\right)^{3}+\cdots~,
\end{align}

Considering the cubic form of these perturbed tetrads and inserting all the solutions from linear order, we find that the contribution
from the cubic terms $f_{TTT}(T^{(0)})\left(T-T^{(0)}\right)^{3}$ is a total derivative in terms of the single variable $\psi$, which takes the
form as
\begin{align}  -\frac{32f_{TTT}\partial_{j}\partial^{j}\psi\left(\partial_{i}\partial^{i}\psi\partial_{k}\partial^{k}\psi+2\partial^{i}\psi\partial_{k}\partial^{k}\partial_{i}\psi\right)}{3a^{6}}\sim0
\end{align}
after performing integration by parts. 
Therefore, without any loss, we can get cubic perturbation Lagrangian 
$\mathcal{L}_{3}$ by inserting the above perturbed tetrads into  
(\ref{eq: EFT for f(T)}). After performing integration by parts, we finally acquire
\begin{align}
 & \!\!\!\!\mathcal{L}_{3}=M_{P}^{2}a^{3}\Big[\!-\frac{f_{T}}{3H^{3}a^{4}}(\partial^{2}\psi)^{2}\dot{\psi}-\frac{2f_{TT}}{3H^{2}a^{6}}\partial^{2}\psi\left(5\partial^{2}\psi\partial^{2}\psi+2\partial_{i}\partial_{j}\psi\partial^{i}\partial^{j}\psi\right) \\ \nonumber
 & \qquad\ \ \ \ \ \ \ \  +\frac{f_{T}}{9H^{2}a^{4}}\psi(-2\partial^{2}\psi\partial^{2}\psi\!+\!5\partial_{i}\partial_{j}\psi\partial^{i}\partial^{j}\psi) +\frac{2\dot{f}_{T}}{3H^{3}a^{4}}\psi(\partial^{2}\psi\partial^{2}\psi\!-\!2\partial_{i}\partial_{j}\psi\partial^{i}\partial^{j}\psi) \\ \nonumber
 & \qquad\ \ \ \ \ \ \ \ -\frac{8\dot{f}_{TT}}{3Ha^{4}}\psi(\partial^{2}\psi\partial^{2}\psi\!-\!\partial_{i}\partial_{j}\psi\partial^{i}\partial^{j}\psi)\Big].\label{fTcubic}
\end{align}
 As we can see, there are many equivalents up to a total derivative, 
namely $\mathcal{L}_3$  above is not the only possible 
form. According to the specific form of $\mathcal{L}_3$, it is clear that no 
propagating scalar mode appears.

\section{The strong coupling scale of $f(T)$ gravity}
\label{sec: strong coupling scale}
Let us now come to the main question of the present manuscript and discuss the strong coupling issue. 
As we showed in the previous subsections, we confirm that there is no propagating scalar mode of $f(T)$ gravity on both linear and second-order perturbation. 
However, considering the existence of the nonvanishing first order time-derivative $\dot{\psi}$ in $\mathcal{L}_{3}$, 
the extra scalar mode may become dynamical in higher order perturbations. 

In the following we give a simple estimation of the strong coupling scale, since it will provide a complementary way to examine the possible strong coupling problem in $f(T)$ 
gravity based on the result in cubic form.  
For simplicity, we employ the notation
\begin{align} 
f_{T}=\frac{3M^{4}}{2H^{2}M_{P}^{2}}~,
\end{align}
and rewrite the single-variable form of quadratic action $\mathcal{L}_{2}$ \eqref{fTk4}as 
\begin{align}
S_{2}\sim-\frac{M^{4}}{2}\int 
dt\,dx^{3}a^{3}\frac{(\partial^{2}\psi)^{2}}{H^{4}a^{4}}~,
\end{align} 
where $M$ can be considered as the cutoff of the EFT applicability. 
The leading term $\frac{(\partial^{2}\psi)^{2}}{H^{4}a^{4}}$ for a given energy scale $E$ yields $\frac{E^{-1}p^{-3}M^{4}\left(p^{4}\psi^{2}\right)}{H^{4}}$, which we assume to be of order one. Then the amplitude of $\psi$ is estimated as
\begin{align}
\psi\sim\frac{H^{2}E^{\frac{1}{2}}}{M^{2}p^{\frac{1}{2}}} .
\label{scalepsi}
\end{align}

In usual perturbative analysis without an indication of strong coupling, the linear perturbation dominates over the nonlinear higher order terms. 
Therefore, we can estimate the energy scale $E_{cubic}$, at which the cubic terms become comparable to the quadratic term in our case. 
Generally, the physical momentum $p$ is related to $E$ through the dispersion relation obtained in quadratic order. 
In the case of $f(T)$ gravity, we can not address it through the dispersion relation since there is no time derivative in the single-variable form of quadratic action. 
Once the scale connection between $E$ and $p$ is given, we can obtain a specific expression of the energy scale and compare it with the cutoff scale $M$. 
For this purpose, we introduce a dimensionless parameter $\beta$ accounting for the ratio $\beta\sim\frac{EH}{p^{2}}$. The specific form of the ratio can be 
obtained from a dispersion relation $\omega^{2}\propto k^{4}$, which has already 
been studied in \cite{Motohashi:2019ymr,DeFelice:2022xvq}. Finally, we mention 
that  we focus on the modes that are deep inside the horizon, namely
$\frac{p^{2}}{H^{2}}\sim\frac{E}{\beta 
H}\gg1$. 

If $\beta>1$ is satisfied, then $(\partial^{2}\psi)^2\dot{\psi}$ is the leading 
term 
in cubic order. A rough estimation of this term is
\begin{align}
\frac{(\partial^{2}\psi)^{2}\dot{\psi}}{H^{3}a^{4}}\sim\psi^{3}EH^{-3}p^{4}\,. 
\end{align}
Hence, in this case  we estimate the scale $E_{cubic}$ by the ratio 
$\mathcal{L}_3/\mathcal{L}_2$ as
\begin{align}
\frac{E}{H}\psi\mid_{E=E_{cubic}}\sim1\rightarrow 
E_{cubic}\sim\left(\frac{M^{3}}{\beta H^{3}}\right)^{\frac{1}{5}}M~.
\end{align}
Higher order terms, such as third-order perturbation terms, may become 
comparable with the linear order term below $E_{cubic}$. Since only cubic order 
terms are considered in this work, $E_{cubic}$ should be regarded as an upper 
bound of the strong coupling scale \cite{Motohashi:2019ymr}. 
Actually, the condition $\frac{E}{\beta H}\gg1$ can be used as a requirement on 
$\beta$ as $\beta\ll\frac{E}{H}$. Applying it in the expression of 
$E_{cubic}$  we extract a lower bound of $E_{cubic}$ as
\begin{align}
E_{cubic}\gg\left(\frac{M}{H}\right)^{\frac{1}{3}}M~.
\end{align}
It is evident that $E_{cubic}$ is expected to be much higher than the cutoff $M$, under the reasonable assumption that $\frac{M}{H}\gg1$ in the EFT formalism. 
In other words, these modes are weakly coupled all the way up to the scale $M$. 

We point out here that the discussion above is based on the definition of the parameter $\beta$, the specific form of which can be treated as an effective one. 
With a specific definition of $\beta$ different terms may dominate for different values of $\beta$. 
Therefore, one could find an appropriate way in which all the relevant modes are free of the strong coupling problem 
below the scale $M$. 
Furthermore, note that in the EFT approach $f(T)$ gravity 
can be regarded as a low-energy effective theory. 
It is possible that the strong coupling problem can be eliminated if other operators are introduced in the EFT action beyond the scale $M$ \cite{Mylova:2022ljr}.

As for the contribution from the higher order EFT operators, we have the following consideration. 
First of all, there is no doubt that higher order operators in the EFT action would not contribute to the background. 
For quadratic action, taking $(T-T^{(0)})^{2}$ term into account would not change our main conclusion in linear perturbation: kinetic term is absent, then a hint of the strong coupling issue is still shown in this case.
Regarding cubic term $(T-T^{(0)})^{3}$ considered on second-order perturbation, the leading contribution from this term is just a total derivative after performing integration by parts at this level. Then we can omit it for simplicity when dealing with perturbation up to second order. 
Besides, the key point in this work is providing a rough estimation of the energy scale $E_{cubic}$, at which the strong coupling occurs, indicated by the ratio $\frac{\mathcal{L}_{3}}{\mathcal{L}_{2}}$ . 
The estimation value of this ratio only depends on the dominant term in both perturbation Lagrangian $\mathcal{L}_{2}$ and $\mathcal{L}_{3}$. 
In other words, this energy scale estimation does not require a complete form of the perturbative action. 
We check that higher order operators would not change both leading terms under the same conditions, and also for the final result on the strong coupling scale. From the above considerations, we conclude that the leading order EFT action of $f(T)$ gravity capture most physical properties of the theories, especially the strong coupling behavior, without these higher order operators.

\section{Conclusions}
\label{sec: conclusion}

In particular classes of modified theories of gravity one may have the case where some extra DoFs may vanish in the background and linear perturbation levels, but can appear at higher orders in perturbation theory. 
This behavior has been shown in $f(T)$ gravity for an extra scalar mode around the Minkowski background, which suggests that $f(T)$ gravity may suffer from the strong coupling problem. 
In this work we investigated further the scalar perturbations and the resulting strong coupling issue of $f(T)$ gravity around a cosmological background, applying the EFT approach as an ideal tool to efficiently deal with perturbations and estimations of the strong coupling scale. 

In order to achieve our goal, we firstly revisited the generalized EFT framework of modified teleparallel gravity, and expanded the perturbative EFT action of $f(T)$ gravity up to quadratic and cubic order.
In addition, we considered both linear and second-order scalar perturbations for $f(T)$ theory. Taking second-order into consideration allows not only to check the existence of new dynamical DoF, but also to measure the strong coupling energy scale based on the ratio of cubic to quadratic perturbation Lagrangians.
We found that no scalar mode is present in both linear and second-order scalar perturbations around a flat cosmological background in $f(T)$ gravity, which may suggest a strong coupling problem. 
Then, by assuming a suitable parameterized relation between the energy scale $E$ and physical momentum $p$, we provided a simple estimation of the strong coupling scale. 
The obtained result reflects that the strong coupling problem can be avoided at least for some modes.

We close this work by commenting that such perturbation behaviors may not inevitably lead to a strong coupling problem,  as long as the relevant scale is comparable to the cutoff scale $M$, which represent the applicability of the theory.
On the other hand, one can consider $f(T,B)$ gravity as a more general case and investigate its dynamics in the same method. 
Since it is already known that $f(R)$ gravity has one extra DoF and does not suffer from a strong coupling problem, dynamics of $f(T,B)$ gravity certainly deserves further investigation. 
And the EFT formalism can be a very powerful tool in investigating a general versions of modified teleparallel theory. 
Towards this direction, we expect that a more complete understanding of the strong coupling issue in teleparallel geometry can be provided in future work.

\acknowledgments
We are grateful to Jose Beltr\'an Jim\'enez, Liming Cao, Xian Gao, Jin Qiao, 
Jackson  Levi Said and Yunlong Zheng for helpful discussions. This work is supported in part by National Key R\&D Program of China (2021YFC2203100), by NSFC (11961131007, 11653002, 12261131497, 12003029), by CAS young interdisciplinary innovation team (JCTD-2022-20), by 111 Project for "Observational and Theoretical Research on Dark Matter and Dark Energy" (B23042), by Fundamental Research Funds for Central Universities, by CSC Innovation Talent Funds, by USTC Fellowship for International Cooperation, by USTC Research Funds of the Double First-Class Initiative, by CAS project for young scientists in basic research (YSBR-006). 
ENS acknowledges participation in the COST Association Action CA18108  ``{\it 
Quantum Gravity Phenomenology in the Multimessenger Approach (QG-MM)}''. 
All numerics were operated on the computer clusters {\it LINDA} \& {\it JUDY} in the particle cosmology group at USTC.


\bibliographystyle{JHEP}
\bibliography{reference}

\end{document}